\documentclass[reprint,
 amsmath,amssymb,
 aps,
]{revtex4-2}
  
\usepackage{amsmath}
\usepackage{makecell} % For multi-line cell content
\usepackage[table]{xcolor} % For table row coloring
\usepackage{enumitem} % For controlling item spacing
\usepackage{comment}
\usepackage{booktabs} % For better table lines
\usepackage{multirow} % For multirow cells
\usepackage{graphicx}% Include figure files
\usepackage{dcolumn}% Align table columns on decimal point
\usepackage{bm}% bold math
\usepackage[normalem]{ulem}

\usepackage[colorlinks=true, linkcolor=blue, citecolor=blue, urlcolor=blue]{hyperref}

\begin{document}

\preprint{APS/123-QED}

\title{First observation and measurement of the\\${}^{198}\text{Hg}$ bosonic transition in an optical lattice clock}% Force line breaks with \\

\author{Clara Zyskind}
\author{Thomas Lauprêtre}
\author{Haosen Shang}
\author{Benjamin Pointard}
\author{Rodolphe Le Targat}
\author{Jérôme Lodewyck}

% \altaffiliation[Also at ]{}%Lines break automatically or can be forced with \\
\author{Sébastien Bize}%
\affiliation{LTE (LNE-OP), Observatoire de Paris-PSL, CNRS, Sorbonne Université, Université de Lille, LNE}


%\affiliation{ Second institution and/or address\\This line break forced% with \\}% 
%\affiliation{Third institution, the second for Charlie Author}%
%\affiliation{%Authors' institution and/or address\\This line break forced with \textbackslash\textbackslash}%


\date{\today}% It is always \today, today,
  
\begin{abstract} 

We report the first observation of the magnetic-field-induced $(5d^{10}6s^2)^{1}\text{S}_0–(5d^{10}6s6p)^{3}\text{P}_0$ transition in a bosonic isotope of mercury, $^{198}\text{Hg}$, realized in an optical lattice clock. We characterize this new isotope, determining key features such as the quadratic Zeeman shift, the probe light shift, and the magic frequency. We also report a first comparison between the $^{198}$Hg optical lattice clock and $^{87}$Sr. In this comparison, the $^{198}$Hg clock has a relative frequency stability of $6\times 10^{-16}/\sqrt{\tau/\text{s}}$ and a total relative systematic uncertainty of $6.9\times 10^{-16}$. This comparison yields the first direct determination of the $^{198}\text{Hg}/^{87}\text{Sr}$ optical frequency ratio: $^{198}\text{Hg}/^{87}\text{Sr}~=~2.629~315~734~684~118~1$, with the same relative uncertainty.

%\czedit{We also present the first direct measurement of the optical frequency ratio $^{198}\text{Hg}/^{87}\text{Sr}$. In We report a fractional frequency instability of $6\times 10^{-16}/\sqrt{\tau/\text{s}}$ and a total systematic uncertainty of $... \times 10^{-16}$, yielding a measured frequency ratio of $^{198}\text{Hg}/^{87}\text{Sr}$ = 2......} 
 
%This paper presents the first observation of the $(5d^{10}6s^2)^{1}\text{S}_0 \rightarrow (5d^{10}6s6p)^{3}\text{P}_0$ magnetic field induced resonance of the mercury bosonic isotope $^{198}\text{Hg}$ in an optical lattice clock. Additionally\sbedit{>remove additionally and improve sentence with some more explanation. Make a second sentence for the ratio measurement}, it provides calculations of the physical parameters for this bosonic isotope and reports the first precision measurement of the $^{198}\text{Hg}/^{87}\text{Sr}$ optical frequency ratio. The study achieved a stability of ... $10^{-16} \sqrt{\tau/\text{s}}$ and an uncertainty of ... $10^{-16}$, resulting in a measured optical frequency ratio of: $^{198}\text{Hg}/^{87}\text{Sr}$ = 2......       

\end{abstract}

\maketitle

%\sbedit{Add a few sentences to introduce the topic more generally}. 
%\sbedit{also, one sentence on interest of Hg, independently of the isotope, in particular, the BBR shift. also mention fundamental physics and metrology citing \cite{Dimarcq2024} and \cite{Margolis2024}} 
%exhibits a  that is weakly allowed due to hyperfine mixing, enabling electric dipole transitions with
%\sbedit{mention and add refs to existing work with 199Hg} 

Recent advances in optical frequency metrology, particularly the development of ultrastable lasers with fractional frequency instability lower than $10^{-16}$ \cite{Hafner2015, Matei2017, Yan2025a}, have enhanced the capability of optical clocks. %Taking the full benefit of such lasers requires the possibility to probe the clock transition for duration exceeding 1 second, 
Fully benefiting from such lasers requires the ability to probe the clock transition for duration exceeding 1 s, which is not accessible to all atomic species considered for optical clocks. Achieving the highest stability and accuracy requires selecting atomic transitions with longer lifetimes, in addition to minimal sensitivity to external perturbations.
    
%\czedit{Recent advances in optical metrology, particularly the development of ultrastable lasers with fractional frequency instability lower than $10^{-16}$ \cite{Hafner2015, Matei2017, Yan2025a}, have enhanced the capability of optical clocks. However, not all atomic species can fully benefit from these technological improvements. Achieving the highest stability and accuracy in optical frequency standards requires carefully selected atomic transitions with longer lifetimes and minimal sensitivity to external perturbations. \textit{(Optional : Precise characterization of new atomic transitions is also essential for the ongoing effort to redefine the SI second \cite{Dimarcq2024, Margolis2024}, to exploit the full potential of optical clocks, and to enable new tests of fundamental physics [ref?].)}}

%Precise measurements of new atomic transitions are essential for meeting the criteria to redefine the SI second \cite{Dimarcq2024, Margolis2024}, unlocking the full potential of optical clocks and enabling applications such as tests of fundamental physics \czedit{[some ref?]}. Achieving the highest stability and accuracy in optical frequency standards requires carefully chosen transitions with long lifetimes, narrow natural linewidths, and minimal sensitivity to external fields.

Among neutral atoms, mercury offers promising properties for optical lattice clocks. Its low sensitivity to thermal radiation and stray electric fields makes it significantly less affected by blackbody radiation than other species \cite{Hachisu2008}. To date, only the $^{1}\text{S}_0–{}^{3}\text{P}_0$ clock transition of fermionic $^{199}\text{Hg}$ has been studied in a lattice clock \cite{Petersen2008a, Hachisu2008, Yamanaka2015, Tyumenev2016}. This transition is weakly allowed by hyperfine mixing, which governs the spontaneous emission rate from $^{3}\text{P}_0$ to $^{1}\text{S}_0$, setting the $^{3}\text{P}_0$ lifetime. In $^{199}\text{Hg}$, the lifetime is short, $\sim1.5$ s \cite{Porsev2017, Bigeon1967}, which may limit the advantages of next-generation ultrastable lasers \cite{Hafner2015, Matei2017} that enable longer probing times and improved stability \cite{Masoudi2015, Oelker2019, Hinkley2013}.

%In $^{199}\text{Hg}$, this lifetime is relatively short—estimated at 1.3 s \cite{Porsev2017} or 1.7 s \cite{Bigeon1967}—compared to the $^{3}\text{P}_0$ lifetime in $^{87}\text{Sr}$ and $^{171}\text{Yb}$. This shorter excited-state lifetime may constrain the benefits of next-generation ultrastable lasers \cite{Hafner2015, Matei2017}, enabling longer probing times for improved measurement stabilities \cite{Masoudi2015, Oelker2019, Hinkley2013}.  
 
\begin{figure}[h]
\centering
\includegraphics[width=1\columnwidth]{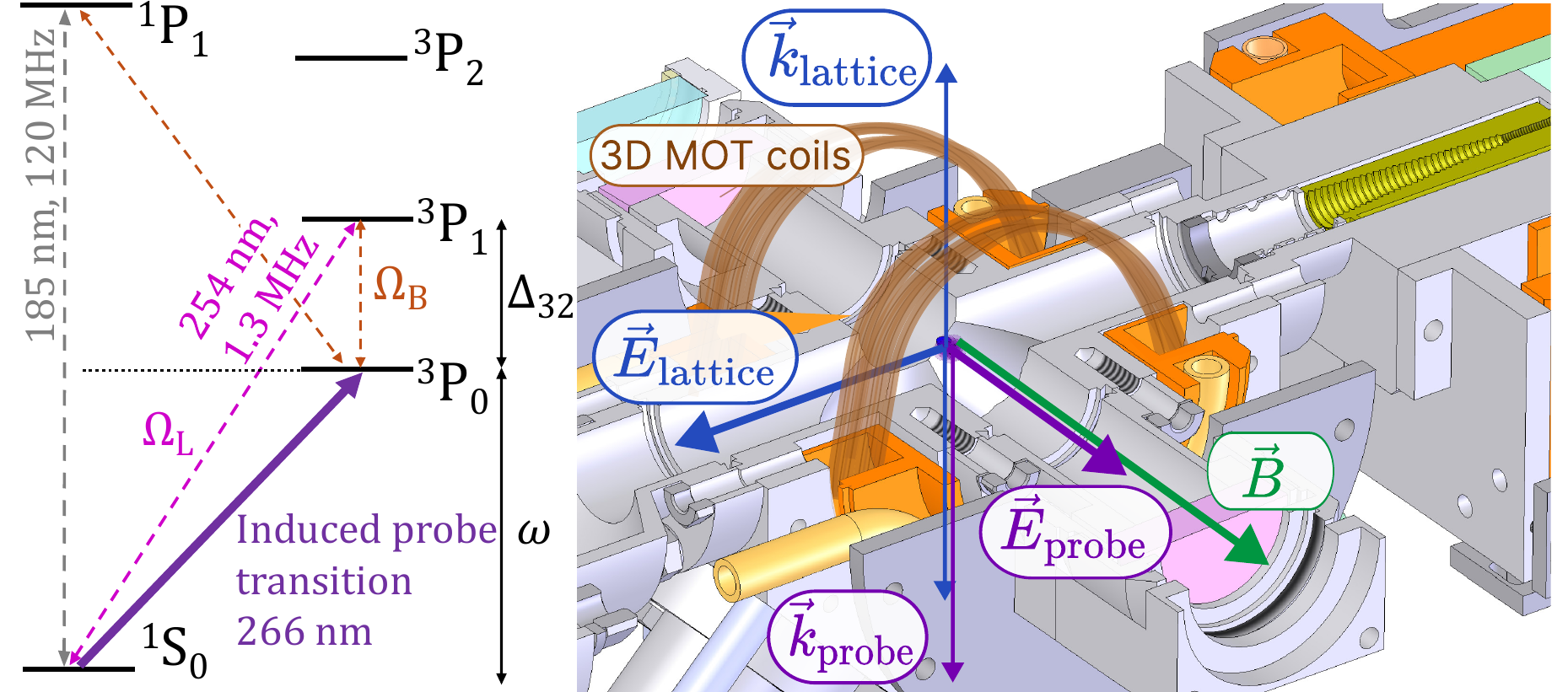}% Here is how to import EPS art
\caption{\label{fig_description_setup} (Left) %\sbedit{We will have see how to optimize this figure. We probabmy want to remove the word 'cooling 254 nm' because it is probably creating confusing here. We should probably solely focus on explaning the probing scheme. Showing the detuning $\Delta_{23}$ could be helpful but I am not sure}
The ${}^{198}\text{Hg}$ bosonic clock transition magnetic-field induced mixing scheme: The doubly forbidden ${}^{1}\text{S}_{0}–{}^{3}\text{P}_{0}$ transition is enabled by a strong static magnetic field $\textbf{B}$, coupling the ${}^{3}\text{P}_{0}$ and ${}^{3}\text{P}_{1}$ states via the matrix element $\Omega_B$ through magnetic dipole interaction (dotted orange lines). The cooling transition ${}^{1}\text{S}_{0}–{}^{3}\text{P}_{1}$ with its electric dipole coupling $\Omega_L$ is indicated by the dotted pink line. (Right) Probing configuration for ${}^{198}\text{Hg}$: The probe beam polarization is aligned with the magnetic field from 3D MOT coils. The vertical lattice is polarized perpendicular to the probe beam.%\sbedit{Texts/labels needs to be made larger.}
}
\end{figure}

Using bosonic mercury isotopes circumvents this limitation. With zero nuclear spin, their $^1\text{S}_0–{}^{3}\text{P}_0$ transition is strictly forbidden, so the $^{3}\text{P}_0$ state does not spontaneously decay. This transition can be weakly induced by applying a strong static magnetic field $\textbf{B}$, as proposed in \cite{Taichenachev2008} and demonstrated in $^{88}\text{Sr}$ \cite{Baillard2007, Takano2017, Origlia2018} and $^{174}\text{Yb}$ \cite{Barber2006}. The magnetic field primarily mixes the ${}^{3}\text{P}_0$ and ${}^{3}\text{P}_1$ states (Fig.~\ref{fig_description_setup}) via the magnetic-dipole coupling $\hbar \Omega_B = \langle {}^{3}\text{P}_{0} |\hat{\mu} \cdot \textbf{B}| {}^{3}\text{P}_{1} \rangle$, with $\hat{\mu}$ the magnetic-dipole operator, perturbing ${}^{3}\text{P}_0$ with a small ${}^{3}\text{P}_1$ admixture. As a result, the otherwise forbidden ${}^{1}\text{S}_0–{}^{3}\text{P}_0$ transition acquires a weak electric-dipole amplitude through the ${}^{1}\text{S}_0–{}^{3}\text{P}_1$ coupling $\hbar \Omega_L = \langle {}^{1}\text{S}_0  | \hat{\textbf{d}} \cdot \textbf{E} | {}^{3}\text{P}_1 \rangle$, driven by an optical field of amplitude $\mathbf{E}$ and frequency $\omega$. Residual coupling to other states, such as ${}^{1}\text{P}_1$, also exists (Fig.~\ref{fig_description_setup}). This magnetic-field-induced mixing allows tuning at will of the transition strength, enabling extended probe times to match the coherence time of the probe laser.

However, direct laser excitation of a bosonic mercury clock transition has not yet been observed. Here, we report the first observation of the ${}^{198}\text{Hg}$ bosonic transition in an optical lattice clock and the first measurement of the ${}^{198}\text{Hg}/{}^{87}\text{Sr}$ optical frequency ratio, enabled by several key experimental advances and a challenging search for a narrow transition across a wide uncertainty range.

%\sbedit{we shall improve this sentence>}Achieving this will be a significant challenge, as it will require substantial experimental advancements and a meticulous search for a narrow transition within a broad uncertainty range. \sbedit{one sentence of give the outline} \sbedit{shift this after (section, first observation)>}The best-known value for the clock transition of a bosonic isotope of mercury was for the ${}^{198}\text{Hg}$ isotope, according to paper \cite{NIST_ref198} (but inconsistent with paper \cite{Saloman}).   
  
%\section{\label{sec:level1}Setup description}

The setup for searching the bosonic clock transition builds on the existing experiment used for probing the ${}^{199}\text{Hg}$ fermionic isotope, as described in \cite{Millo2009, Dawkins2010, Mejri2011, McFerran2014, Tyumenev2016, Guo2023}.% \sout{and references therein.} 

%First, we need to generate a well-controlled and strong magnetic field pulse for the magnetic field induced mixing.
First, a well-controlled and sufficiently strong magnetic field is required to induce the mixing. For this, we generate a pulsed magnetic field by using the 3D MOT coils and switching electronics to rapidly reverse the current in one of the coils during interrogation, following the approach used in \cite{Barber2006}. A maximum field of $B = 18.6$~mT can be reached with 12~A current, with rise/fall times as fast as 3/2.8 ms to minimize unwanted dead times. %with fast ramps (rise/fall times 3/2.8 ms) to minimize unwanted dead times. 
Three additional pairs of coils compensate residual background magnetic fields. Their current is set by minimizing the Zeeman splitting of the $^{199}$Hg clock transition.
%adjusted
%Our setup reaches a maximum field of $B = 18.6$~mT with 12~A current

Next, we extended our ultrastable light generation to provide a sufficiently broad continuous tuning range to reach the bosonic isotope. We implemented a setup with an intermediate laser \cite{Clara_PhD}, which is offset phase-locked to the ultrastable light. This offers greater flexibility than the original design \cite{Dawkins2010, Millo2009} and enables ultrastable probing at 266~nm across all mercury isotopes, rather than being limited to ${}^{199}\text{Hg}$. We also ensured that spectral purity was maintained at the present $4\times10^{-16}$ flicker frequency noise level for the ultrastable light \cite{McFerran2012} and that the system remains compatible with future improvements well beyond this stability.  %Next, we extended our ultrastable light generation to provide a sufficiently broad continuous tuning range to reach the bosonic isotope, while preserving spectral quality now (flicker noise of $4\times 10^{-16}$ \cite{McFerran2012}) and beyond. 

In order to facilitate the search for the transition, achieving the highest possible coupling in our setup was critical. As it scales with intensity, we first enhanced our deep-UV probing light to a maximum of 400 µW at 266~nm. By then reducing the probe waist to 63.5 µm, intensities as high as $I_0 = 6.3~\text{W/cm}^2$ could be reached at the center of the beam.
%By optimizing the probe waist to 63.5 µm, we obtained an intensity at the center of the beam of $I_0 = 6.3~\text{W/cm}^2$ at the atoms. 
%Lastly, because the coupling increases with intensity, we enhanced our deep-UV ultrastable light to reach a maximum of 400 µW at 266 nm. Optimizing the probe waist to 63.5 µm yielded an intensity of $I = 6.32~\text{W/cm}^2$ at the atoms.   
With the probe polarization aligned to the applied magnetic field $B$ (Fig.~\ref{fig_description_setup}), the estimated Rabi frequency is $\Omega_{12} = \frac{\Omega_L \Omega_B}{\Delta_{32}} = \alpha \sqrt{I} B$ \cite{Taichenachev2008}, giving $\Omega_{12}/2\pi \simeq 75$ Hz, where $\alpha$ is the magnetic-field-induced coupling coefficient and $\Delta_{32}$ the angular frequency detuning between ${}^3\text{P}_1$ and ${}^3\text{P}_0$.

%The best-known bosonic mercury clock frequency corresponded to ${}^{198}\text{Hg}$ \cite{NIST_ref198, Saloman}, motivating our focus on this isotope to reduce the difficulty of searching for the transition. We searched for the ${}^{198}\text{Hg}$ bosonic transition directly on atoms trapped in the lattice. A potential challenge was the unknown exact lattice magic frequency for ${}^{198}\text{Hg}$, as a large detuning from it could distort the resonance. To mitigate this, we set the lattice to the known magic frequency of fermionic ${}^{199}\text{Hg}$ \cite{Yamanaka2015, Tyumenev2016}, as calculations, confirmed by tests with this fermionic isotope, indicated the difference would be small enough to preserve good contrast. For the search, the trap depth was set to $92.4~E_r$, with the recoil energy $E_r/h = 7.57$ kHz.

With such a weak coupling, a frequency range for the search that is too large becomes prohibitive.
The best-predicted bosonic mercury clock transition corresponded to ${}^{198}\text{Hg}$ \cite{NIST_ref198, Saloman},  motivating our focus on this specific isotope. %\sout{When looking for the transition in} 
As ${}^{198}\text{Hg}$ atoms are trapped in the optical lattice, another potential challenge was the lack of knowledge on the exact lattice magic frequency \cite{Katori2003}. %\sout{This is because}
Indeed, a largely detuned lattice frequency broadens and degrades the contrast of the clock resonance.
We set the lattice frequency to the magic frequency of the fermionic ${}^{199}\text{Hg}$, as calculations and experimental tests with this isotope showed that contrast would be preserved over the expected lattice detuning range. The lattice trap depth was set to $92.4~E_r$, with the recoil energy $E_r= h \times 7.57$ kHz.

%\sbedit{somewhere in this paragraph, it would be logical to give the value of the trap depth} We chose to search for the ${}^{198}\text{Hg}$ bosonic transition directly on atoms trapped in the lattice. This approach posed a potential challenge, as the exact value of the ${}^{198}\text{Hg}$ magic frequency was unknown, and a large detuning from it could distort the resonance, making the transition difficult to observe \cite{Yi2011}\sbedit{>change or remove reference}. To mitigate this, the lattice frequency was set to the magic frequency of the $^{199}$Hg fermionic isotope, as simulations\sbedit{>modify. simulation might not be the best term here} indicated that any difference between the fermionic and bosonic magic frequencies would likely be small enough to have no significant impact on contrast.\sbedit{>in general, let try to make ths paragraph a bit shorter}

%The MOT for bosonic isotopes was obtained using the existing setup \cite{Mejri2011, Guo2023}, by tuning the MOT laser to the ${}^{1}\text{S}_0 \rightarrow {}^{3}\text{P}_1$ transition of ${}^{198}\text{Hg}$ \cite{Witkowski2019}. The measured MOT loading time constant was 1.2 s, and the lattice lifetime 200 ms, both comparable to those of fermionic ${}^{199}\text{Hg}$.
When trapping  ${}^{198}\text{Hg}$, we measured a MOT loading time constant of 1.2 s and a lattice lifetime of 200 ms, both comparable to those of fermionic ${}^{199}\text{Hg}$.

%The MOT laser frequency was detuned to 1181.555932 THz, aligning with the $^{1}\text{S}_0 \rightarrow {}^{3}\text{P}_1$ transition value from \cite{Witkowski2019}. This frequency is stabilized to the saturated absorption feature observed in a mercury vapor cell \cite{Petersen2008a}. The measured MOT lifetime for the bosonic isotope was about 1.2 s, and the lattice lifetime was approximately 250 ms, both comparable to those observed for the ${}^{199}\text{Hg}$ fermionic isotope.

%\sbedit{start by saying that we chose to search for the transition directly on atoms trapped in the lattice. Then explain why it is potentially a problem since we do not know the exact value of the magic wavelength} To search for the narrow clock transition, we started with the estimated frequency value from \cite{NIST_ref198}. The lattice frequency was set to the magic frequency of the fermionic isotope, as simulations indicated that any detuning difference between fermionic and bosonic isotopes would likely be small enough to have no impact on contrast. 

\begin{figure}[h]
    \centering
    % Haystack figure on top
    \includegraphics[width=1\columnwidth]{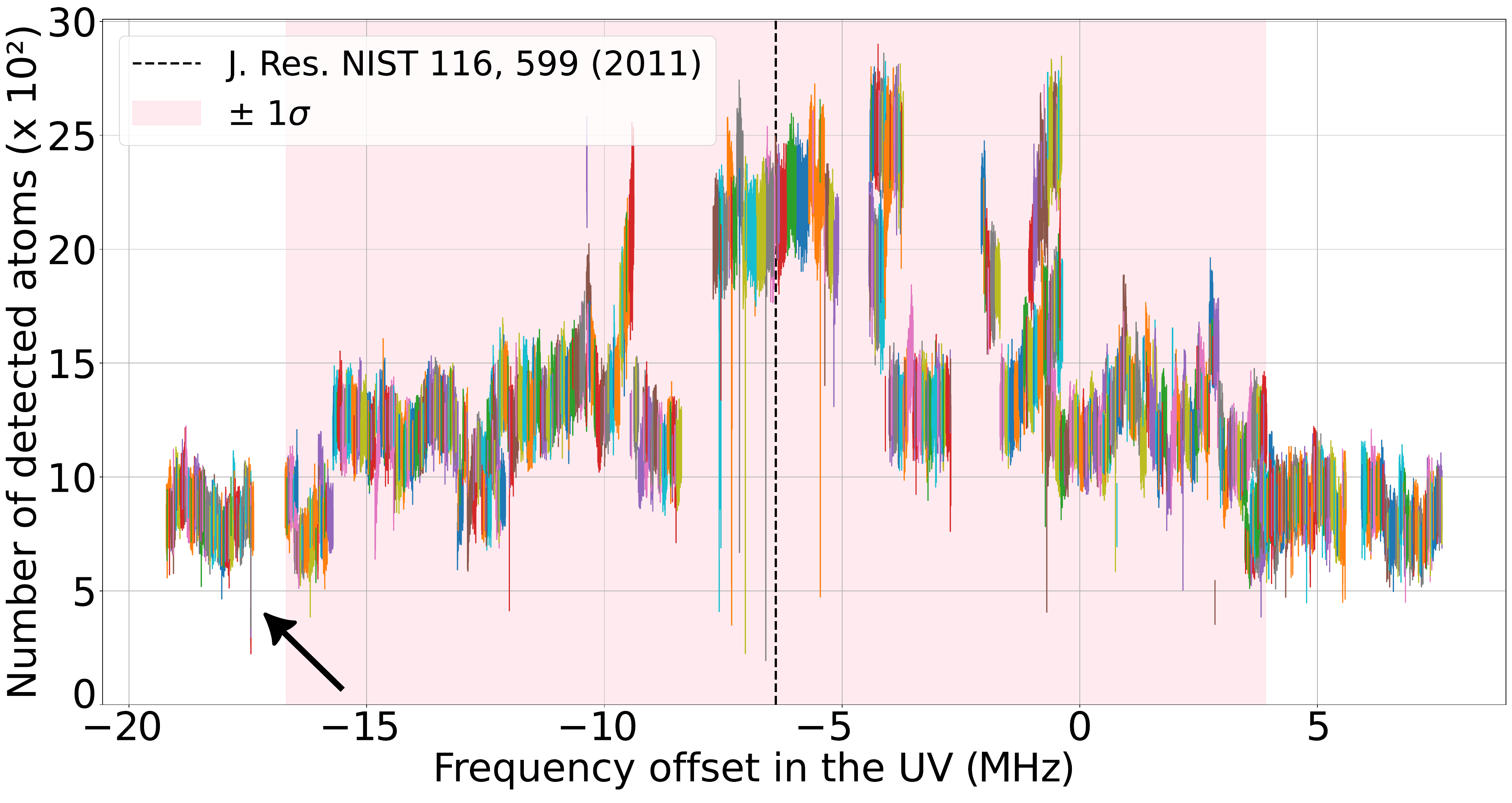}
    \vspace{0.4cm} % Optional: 
    \begin{minipage}{0.482\textwidth}
        \centering
        \includegraphics[width=\textwidth]{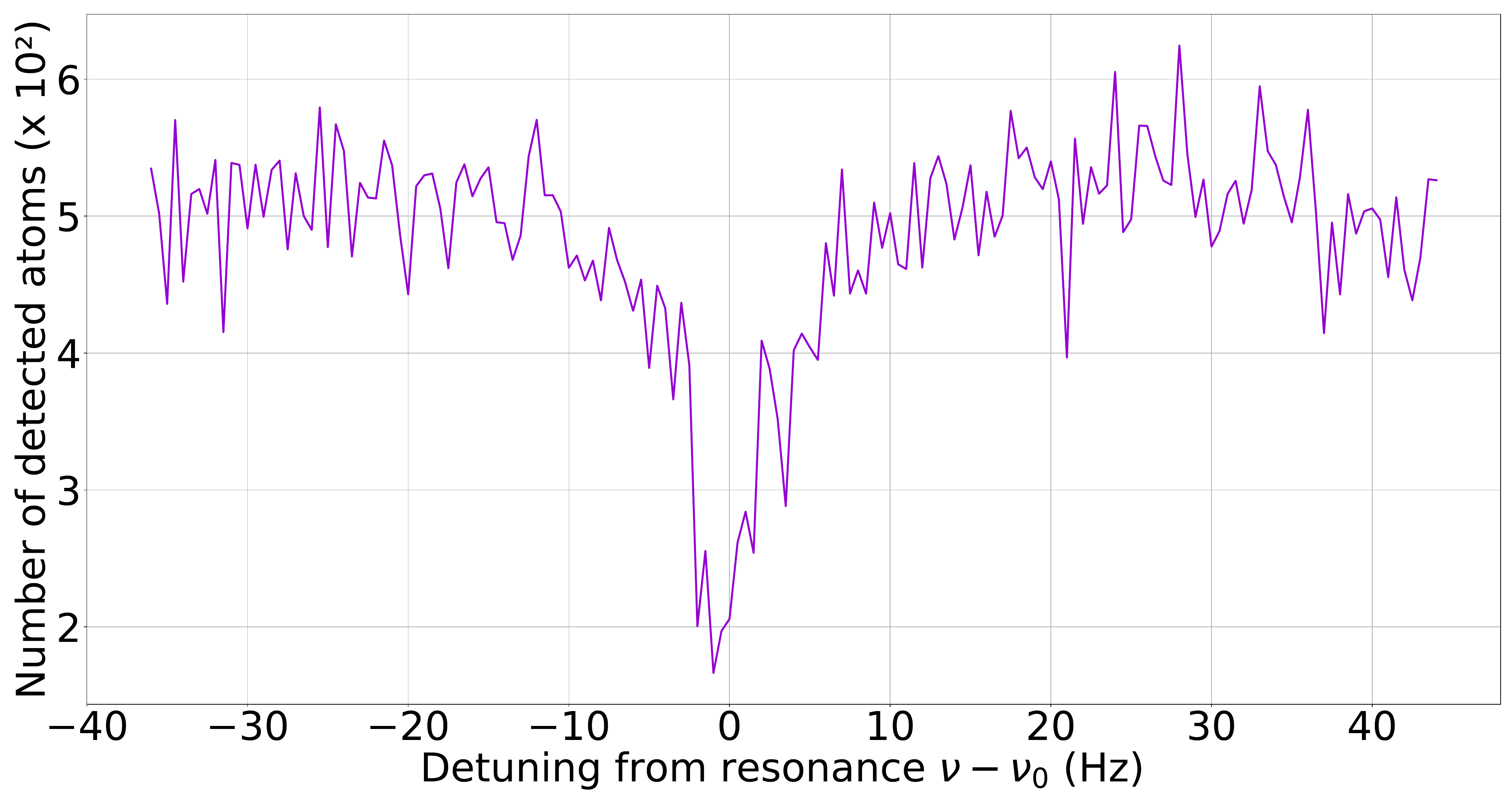}
    \end{minipage}
    \vspace{-0.8cm} 
    \caption{\label{fig_haystack}(Top) All spectra acquired for the search of the ${}^{198}\text{Hg}$ transition in the optical lattice. Each color corresponds to a scan covering a specific range of 100 kHz by steps of $120$~Hz. 
    %\tledit{\textit{(Thomas comprend pas par rapport aux 120 Hz)}}. 
    In total, more than 25~MHz were scanned before successfully identifying the transition (black arrow).
    %\tledit{The variations on the detected level come from variations of the available MOT laser power and don't indicate anything physical.}
    (Bottom) Narrowest ${}^{198}\text{Hg}$ spectroscopy obtained, after optimization: for 200~ms probe time, the full width at half maximum is 4~Hz.}
\end{figure}
    
%Using the experimental setup and all its improvements above, we thoroughly explored the uncertainty range
Once the experimental setup ready, we conducted a thorough exploration over the predicted frequency range for the ${}^{198}\text{Hg}$ bosonic transition \cite{NIST_ref198}, scanning a total range of 25 MHz by 120 Hz steps (Fig.~\ref{fig_haystack}, top). This meticulous search led to the first successful observation of the bosonic transition. Its frequency lies slightly beyond the $\pm 1\sigma$ uncertainty of \cite{NIST_ref198} and beyond $\pm 2\sigma$ of \cite{Saloman}.

%initial measured
Once the transition found, we measured an experimental Rabi frequency to actually be $\Omega_{12}/2\pi~=~45$~Hz.
%Reducing the applied magnetic field to 1.86 mT (1/10$^{\text{th}}$ of its maximum value) narrowed the transition linewidth, allowing a 200 ms probe time and a 4 Hz full-width at half-maximum 
Optimizing parameters (magnetic field reduced by a factor 10 and a lower probing power) allowed a probing time of $200$~ms giving spectra with a $4$~Hz full width at half maximum (Fig.~\ref{fig_haystack}, bottom). 
Operating a clock sequence with such parameters and analyzing the noise of frequency corrections, we estimate that the clock stability surpasses by a factor $\sim 2$ the stability previously achieved with the fermionic isotope \cite{Guo2023}. 
From comparisons against local Sr clocks, we infer stabilities of $\sim 6\times 10^{-16}/\sqrt{\tau/s}$ for the ${}^{198}\text{Hg}$ bosonic clock. 
%\czedit{Under these conditions, the estimated stability of the bosonic mercury clock is approximately $4\times10^{-16} \sqrt{\tau/s}$, obtained by scaling the single-cycle short-term stability with a factor dependent on the loop gain parameters. This stability already surpasses by a factor of two the stability previously achieved with our fermionic isotope clock setup [à revoir]}. 
%\begin{figure}[h]
%\centering
%\includegraphics[width=1\columnwidth]{haystack_and_zoom_in_with_0_cropped.pdf}% Here is how to import EPS art
%\caption{\label{fig_haystack} \sbedit{Let see how/where to show the optimized spectrum too.} All spectra acquired in the optical lattice before finding the ${}^{198}\text{Hg}$ bosonic transition. Each color corresponds to a different spectrum covering a specific range of $\sim$ 100 kHz. In total, more than 25 MHz were scanned in the lattice before successfully identifying the bosonic transition. The enclosed upper left spectroscopy is a zoom in on the first observation of the bosonic transition in the optical lattice.}
%\end{figure}

\begin{comment}
    \begin{itemize}[noitemsep]
    \item[-] Bosonic MOT
    \item[-] Bosonic lattice
    \item[-] Research in the lattice
    \item[-] Uncertainty litterature
    \item[-] Found needle in haystack
    \item[-] Best parameters for now, 4 Hz FWHM and 200 ms probe
    \item[-] Estimated stability Hg alone
\end{itemize}
\end{comment}

%\section{Study of the ${}^{198}\text{Hg}$ bosonic transition}

%\subsection{Quadratic Zeeman shift}

Having identified the new transition, we measured key systematic effects to characterize its properties. First, a high magnetic field is required to couple the transition. Although there is no linear Zeeman shift due to zero nuclear spin, a significant quadratic Zeeman shift (QZS) is expected and observed, necessitating accurate measurement. The QZS is given by $\Delta_B = \beta B^2$, with $\beta/2\pi$ in Hz/T$^2$. We measured $\beta$ using an intertwined differential method: two alternating configurations, differing only in the applied magnetic field, produce different QZS from which we extract a differential shift value.
%We measured $\beta$ using an intertwined differential method: \sbedit{<check this sentence>} two alternating configurations where only the applied magnetic field is varied yield different QZS values, and the differential shift between them is extracted.

%\sbedit{clarify the description of an intertwined differential measurement as well as possible here. And then try to avoid repetitions elsewhere.} 
%The coefficient $\beta$ was measured using an intertwined differential method, involving two alternating configurations in which the induced magnetic field was the only varied parameter. Each configuration has a different QZS and the differential shift between them is then deduced.

\begin{figure}[h]
\centering
\includegraphics[width=1\columnwidth]{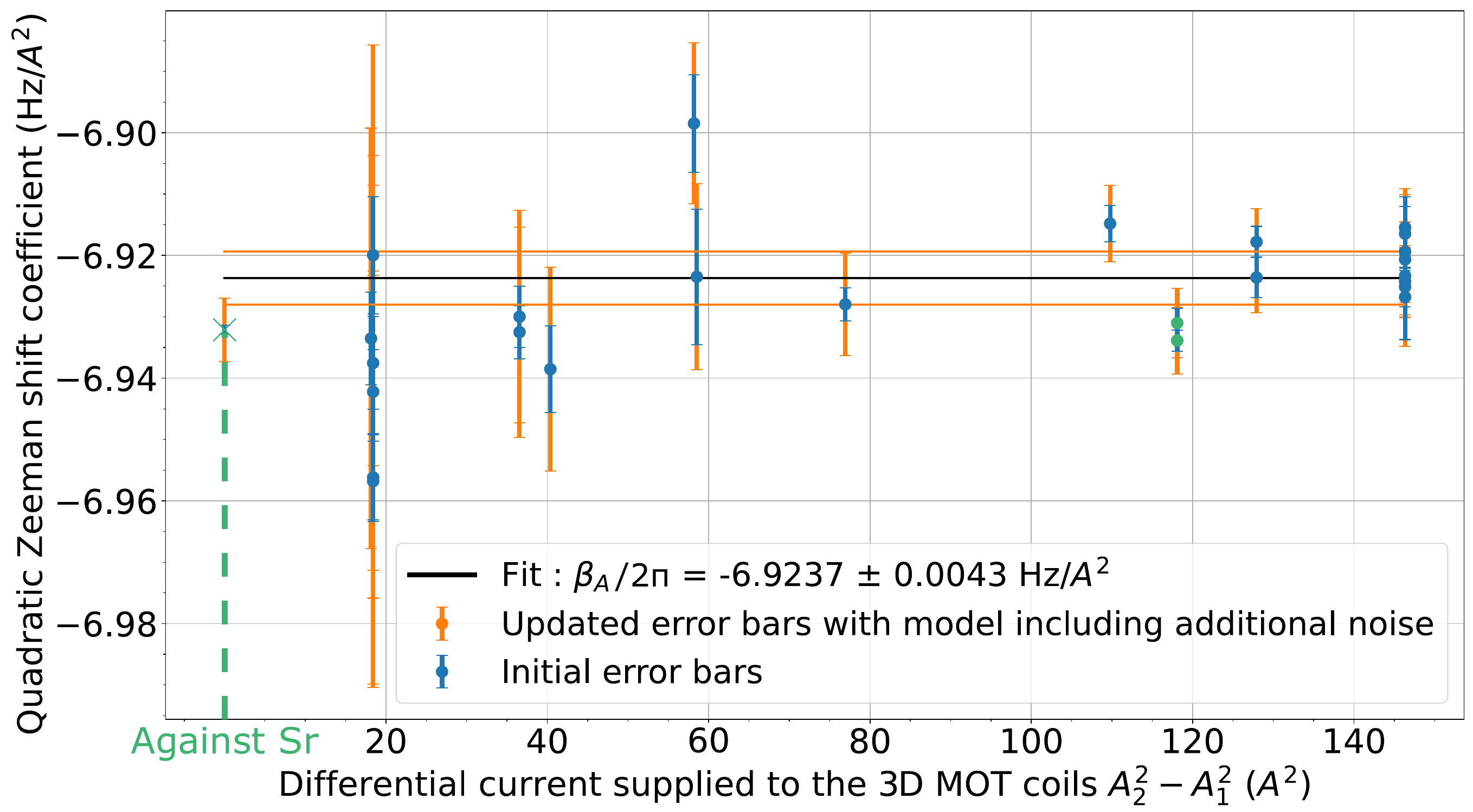}% Here is how to import EPS art
\vspace{-0.3cm}
\caption{\label{fig_QZS}Measurement of the quadratic Zeeman shift coefficient $\beta_{\text{A}}/2\pi$ with initial error bars (blue) and error bars incorporating a model with additional uncorrelated noise (orange). See text for more details. Blue and green dots represent differential measurements from Mar.-Apr. 2024 and Sep. 2025, respectively. 
%Cross symbols are measurements made in September 2025. Two of these are differential measurements similar to the previous ones. 
The cross symbol is a value deduced from 3 direct measurements against a Sr clock at currents of 1.1 A, 3.6 A and 10.9~A \cite{note_1}. The black line represents the estimated coefficient returned by the fit of the measurement values with the modeled error bars. The orange lines represents our final uncertainty range ($\pm1\sigma$), where $\sigma$ is defined here as the smallest modeled error bar.}
%\sbedit{adapt/improve this sentence}We decide that a reasonable uncertainty is a conservative uncertainty estimation, where the uncertainty is defined as the smallest corrected error bar.}
\end{figure}

%To achieve a more precise measurement of the high QZS, we delved deeper by conducting a detailed analysis \sbedit{here try to give a better sense (non-linearity of control, transients, long term reproducibility)} of the current supplied to the 3D MOT coils responsible for generating the induced magnetic field, as outlined in \ref{SupplementaryMaterialII}.
To achieve the required accuracy in determining $\beta$, we carefully characterized the current source generating the applied magnetic field. Nonlinearities in current control were accounted for, transients from rapid switching were mitigated via optimized timing, and long-term reproducibility was verified through current monitoring (yielding a 10 µA uncertainty). In Fig.~\ref{fig_QZS}, we show the measured QZS coefficient and its statistical uncertainty (blue points and error bars) against the difference of squared currents, where $\beta_{\text{A}}/2\pi$ is expressed in units of Hz/A$^2$.
%\sout{Points are plotted against the difference of squared currents.} 
Most data points were taken during an initial round, supplemented by 3 points measured 20 months later. Their average gives the mean value and uncertainty. However, residual inconsistencies in the QZS data were observed in this initial dataset, with the two most inconsistent points differing by $6.3\sigma$. To obtain a reasonable uncertainty despite the unknown origin of these fluctuations, we modeled the dispersion by adding uncorrelated noise and adjusted the error bars accordingly to reflect the total noise \footnote{The additional uncorrelated frequency noise was estimated to be 0.61 Hz, corresponding to the value required to bring the goodness of fit to the nominal value of 0.5, or equivalently, the Birge ratio to 1.}. The updated results, shown in Fig.~\ref{fig_QZS} (orange error bars), give a QZS coefficient of $\beta_{\text{A}}/2\pi = -6.9237 \pm 0.0043$~Hz/A$^2$. For the lowest value of the current we used so far ($1.13$~A), the corresponding fractional uncertainty on the QZS correction is $4\times 10^{-18}$.

%We also need to take into account for uncertainties arising from magnetic field fluctuations and residual magnetic field. 
Uncertainties arising from magnetic field fluctuations and residual magnetic field also need to be addressed.
With a magnetic field probe, we detect fluctuations of 1~µT. The residual field is minimized to less than 1.5 µT. %\tledit{\sout{For the same lowest current value, it corresponds to uncertainties at worst $9 \times 10^{-18}$ and $1.4\times 10^{-17}$.}} \czedit{Pas dacc ici}
These fluctuations were estimated to have a negligible impact on the measurement of $\beta$. In Table \ref{table_198_budget}, we give the QZS correction and uncertainties for $3.6$~A, the highest value applied to our frequency ratio determination. 
%in our uncertainty budget (Table \ref{table_198_budget}).} Using a magnetic field probe, we measured fluctuations of ~1 µT, corresponding to a fractional frequency uncertainty of about $8 \times 10^{-18}$. The residual magnetic field was canceled to 1.5 µT, giving a potential QZS uncertainty of $3.8 \times 10^{-18}$\sbedit{? $1.2\times 10^{-17}$}, also included in the budget. These fluctuations were estimated to have a negligible contribution for the measurement of $\beta$.
%\sbedit{we have to discuss how this is estimated (it is seems high to me) and whether a contribution of this to the measurement of $\beta$ should be taken in into account}.

Lastly, to compare the QZS coefficient with its theoretical value (Table \ref{fig_table_rabi}), it was converted to magnetic field units using a geometric factor of $1.527 \pm 0.04$~mT per ampere of current flowing through the 3D MOT coils. %\sout{The results are summarized in Table \ref{fig_table_rabi}.}

%\subsection{magnetic field induced coefficient}
Another essential parameter of the bosonic isotope probing scheme is the magnetic-field-induced coupling
coefficient $\alpha$, describing the coupling between ground and excited states via the external applied magnetic field. To determine $\alpha$, we measured Rabi frequencies at constant probe intensity $I$ while varying $B$, yielding a linear relationship whose slope gives $\alpha$. We accounted for experimental factors affecting the measured Rabi frequencies to ensure accuracy. The 63.5~µm probe waist is comparable to the atomic distribution size, causing the atoms to experience an inhomogeneous probe beam intensity and hence different Rabi frequencies, leading to damped Rabi oscillations.
%\tledit{Using a simplified model that averages the Rabi oscillations over the transverse atomic distribution, we estimate a ratio between the time when the Rabi flopping reaches its first maximum and the Rabi half-period in the ideal case of an atom at the center of the probe beam (thereby seing the peak intensity). 
We used a simplified model that averages the Rabi oscillation over the transverse atomic distribution to estimate a factor linking the first maximum of the measured Rabi flopping curve to the Rabi half-period in the ideal case of an atom at the center of the probe beam, thereby seeing the peak intensity.
%We find that the correction factor to apply is 1.5
We find a correction factor of 1.3, in line with strong inhomogeneity. The resulting value of $\alpha$ is given in Table~\ref{fig_table_rabi}.

%Using a simplified 2D transverse atomic distribution model \sbedit{improve model naming} \czedit{(je vois toujours pas trop quoi mettre ici...)}, we corrected the measured Rabi times by a factor of 1.5 to approximate the ideal Rabi time for a perfectly homogeneous probe beam. The resulting $\alpha$ is given in Table~\ref{fig_table_rabi}.
%We did a very simplified model adding a transverse 2D atomic distribution to consider this effect. As a consequence, the measured Rabi times were adjusted by a factor of 1.5 to approximate the ideal Rabi time for a perfectly homogeneous probe beam. The result obtained for the calculation of $\alpha$ including this effect is provided in Table~\ref{fig_table_rabi}.  

%\subsection{Magic frequency}

%\sbedit{Improve this sentence and before it, add introductory sentence reminding that the magic frequency an atomic property specific to each species or isotope that is essential to measure accurately in the context of lattice clocks. Also, from a purely semantic point of view we may try to use non perturbing frequency and also remind in a short sentence plus citation, what this non-perturbing trap frequency is.}

\begin{figure}[h]
\centering
\includegraphics[width=1\columnwidth]{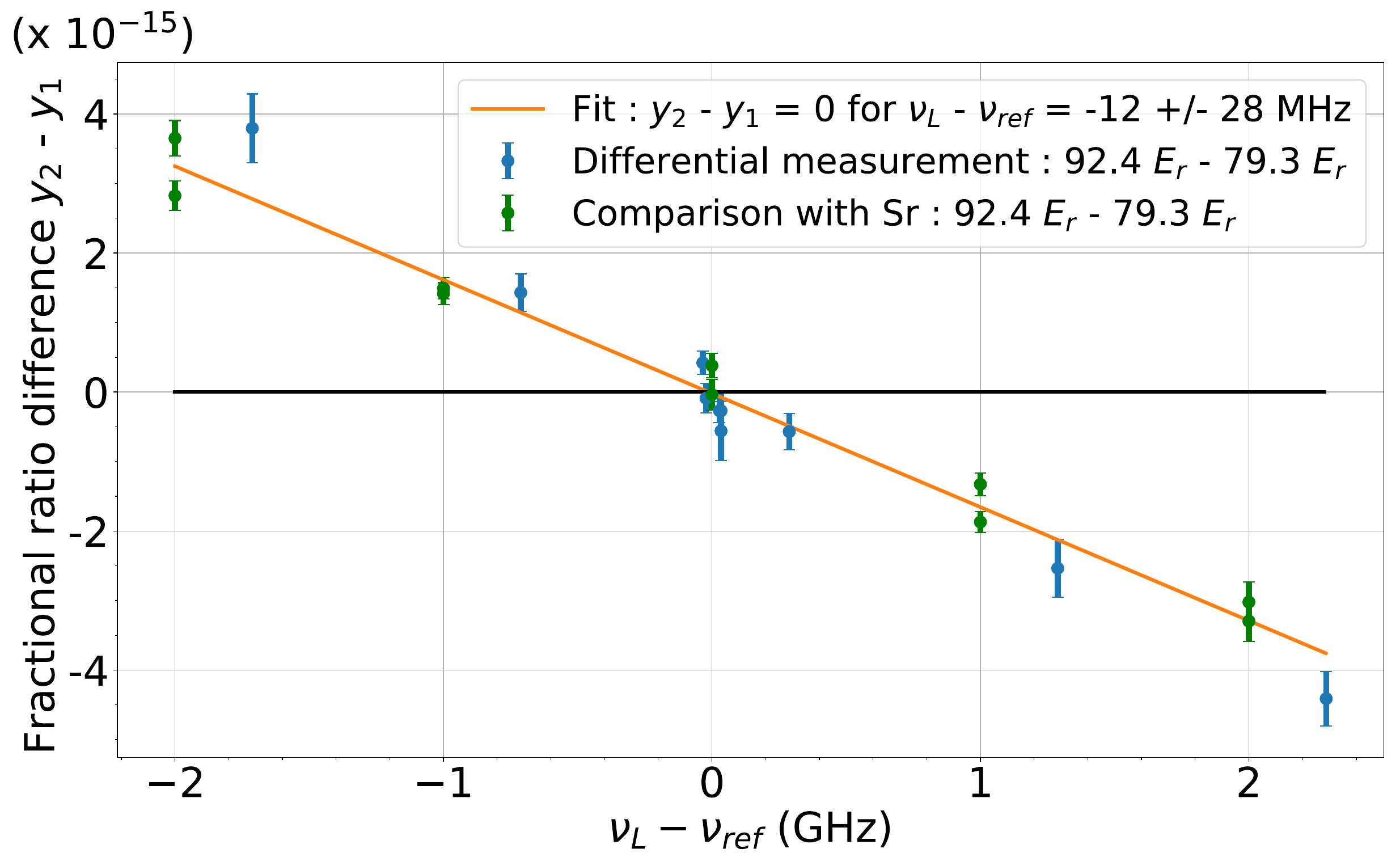}% Here is how to import EPS art
\vspace{-0.4cm}
\caption{\label{fig_magic} Magic frequency determination from two series of differential measurements: one with the Hg clock alone and one assisted by a Sr clock 
\cite{Lodewyck2016, SrClock} to improve stability. %The linear lattice light shift coefficient can be directly derived from the slope of the affine fit of the orange curve. 
The reference used for the lattice frequency is $\nu_{\text{ref}}=826~855~240$~MHz. 
The linear fit (orange) returns the value of the magic frequency $\nu_{L}^{m}=826~855~228\pm 28~\text{MHz}$ for $^{198}$Hg. It also gives the experimental slope of $\eta=(-1.41\pm0.04)\times10^{-4}\text{~Hz}/E_r/\text{MHz}$.  Both are used to estimate the lattice light shift correction and its uncertainty.
%The magic frequency derived from the fit (orange) is $\nu_{L}^{m}=826~855~228\pm 28~\text{MHz}$. The fit also gives the experimental slope of $\eta=(-1.41\pm 0.04)\times10^{-4}\text{~Hz}/E_r/\text{MHz}$, that we use as well 
%\tledit{The linear fit (orange) returns the value of the magic frequency $\nu_{L}^{m}=826~855~228\pm 28~\text{MHz}$ and the experimental slope  $\eta=(-1.41\pm 0.04)\times10^{-4}\text{~Hz}/E_r/\text{MHz}$, which we both use to estimate the lattice light shift correction along with its uncertainty.}
%\sbedit{add comment on the slope.}
}
\end{figure}

Another crucial parameter to measure for an optical lattice clock is the magic frequency, the non-perturbing lattice frequency at which the light shift is canceled at any trapping depth \cite{Katori2003}. The magic frequency is an intrinsic atomic property that varies between species and isotopes, therefore its precise value must be determined for the ${}^{198}\text{Hg}$ bosonic isotope.
%To measure the magic lattice frequency and the lattice light shift for the ${}^{198}\text{Hg}$ bosonic isotope, we performed a series of measurements illustrated in Figure~\ref{fig_magic}. These included a differential measurement method, where the trapping depth was varied, as well as direct measurements against the SYRTE-Sr clock.
To measure the ${}^{198}\text{Hg}$ magic lattice frequency, we performed differential measurements between two different lattice depths: $92.4~E_r$ and $79.3~E_r$. %\tledit{The difference of the transition shift between $92.4~E_r$ and $79.3~E_r$ was measured for different lattice frequencies $\nu_L$ (Fig.~\ref{fig_magic}) 
Fig.~\ref{fig_magic} shows the differential clock shift as a function of the lattice frequency $\nu_L$.
%(x-axis), two depths ($92.4~E_r$ and $79.3~E_r$) were used to measure the transition detuning between them (y-axis). 
The lattice frequency was precisely known and stabilized using an optical frequency comb with the setup described in \cite{Guo2020}. 
%\sbedit{Improve/clarify what is meant exactly. Maybe giving the equation could simplify. approximation justified by frequency resolution and trap depth explored small} 

Given our limited trapping depth range and our frequency resolution, a linear approximation of the lattice light shift is justified to analyze our measurements. In this approximation, the lattice light shift (LLS) is proportional to both the lattice depth (in units of recoil energy $E_r$) and the detuning of the lattice laser from the magic frequency. For a configuration with lattice depth $U_{x}$ ($x$ = 1, 2), the resulting fractional frequency shift is $y_x = \eta (\nu_L - \nu_L^m) U_{x}$ where $\eta$ is the experimental slope of the linear LLS. 
%\cite{Tyumenev2016, Katori2015a}. 
The magic frequency $\nu_L^m$ is found at the detuning where the differential lattice light shift is equal to zero. From this analysis, we extract a magic frequency of $\nu_L^m = 826~855~228 \pm 28~\text{MHz}$, as shown in Fig.~\ref{fig_magic}. 
We note that the corresponding uncertainty (Table~\ref{table_198_budget}) is set by limitations of our setting and not by any fundamental limitation associated to $^{198}$Hg.

%\subsection{Cold collisional shift} 
 
We also studied the cold collisional shift (CCS) which arises from interactions between cold atoms and depends on the atomic density, closely linked to the atom number during interrogation $N_{\text{atoms}}$.
%\czedit{Experimentally, the quantitative characterisation of the imaging detection system gives the number of detected atoms. (pas convaincue de l'info de cette phrase, je préférais avant)} 
To estimate it, we compute the integral of the decay of the atom number inside the lattice over the probe pulse timing, based on the number of atoms detected later in the detection. 
%A key experimental parameter for characterizing it is the atom number during interrogation. To estimate this number, $N_{\text{atoms}}$, we take into account finite lifetime of atoms trapped in the lattice and compute the integral of this decay over the probe duration, based on the number of atoms detected later in the detection. 
%\tledit{ We estimate $N_{\text{atoms}}$ from the imaging detection characterization and by computing the integral of the decay of the atom number inside the lattice over the exact probe duration timing. \textit{(not a fan either)}}
The maximum effective atom number during interrogation is $\sim 10^3$, corresponding approximately to one atom per lattice site at the trap center.
%we modeled the atom count as an exponential decay characterized by the lattice lifetime and calculated the integral of this decay over the probe duration. 
Atom number was varied by turning on the MOT beams later in the sequence while keeping other key parameters unchanged, such as cycle time and MOT loading time, ensuring a method that is as differential as possible. For differential atom numbers up to 550, we observed an absence of shift well within our typical resolution of $10^{-16}$. In these conditions, it is reasonable to analyze these measurements under the simplifying assumption that the fractional CCS is proportional to $N_{\text{atoms}}$. Doing so, the CCS found is $(-3 \pm 14) \times 10^{-20} \times N_{\text{atoms}}$.
%Within our resolution, no clear dependence of the shift on atom number was observed, as the measured shifts were consistent with zero within error bars. 
%\czedit{Therefore,} the fractional CCS is assumed proportional to the atom number $N_{\text{atoms}}$, with proportionality constant $K_{\text{CCS}}$. From these measurements, $K_{\text{CCS}} = (-3 \pm 14) \times 10^{-20}$/atom. 
%The fractional CCS is assumed proportional to the atom number $N_{\text{atoms}}$, with proportionality constant $K_{\text{CCS}}$. To estimate $N_{\text{atoms}}$ during interrogation, we modeled the atom count as an exponential decay characterized by the lattice lifetime and calculated the integral of this decay over the probe period. Atom numbers were varied by turning on the MOT beams later in the sequence while keeping other key parameters unchanged, such as cycle time and MOT loading time, ensuring a method that is as differential as possible. From these measurements, $K_{\text{CCS}} = (-3 \pm 14) \times 10^{-20}$/atom. The effective atom number during interrogation is $\sim 10^3$, corresponding to roughly\sbedit{>approximately} one atom per lattice site at the trap center.
%\sbedit{We could give the estimation of the number of atoms (effective during the probe phase), number of atoms per lattice site and/or the peak average density}

%\subsection{Probe light shift}
The probe light itself induces an AC Stark shift of the clock levels resulting in a shift of the transition frequency proportional to the probe intensity $\Delta_L\text{(rad/s)} = \kappa I$.
%We also define the probe light shift $\Delta_L$ (in rad/s) as linearly dependent on the probe intensity $I$, $\Delta_L = \kappa I$, with $\kappa$ the proportionality coefficient.
To determine $\kappa$, we measured frequency shifts against Sr clocks \cite{Lodewyck2016, SrClock} while varying the probe power $P$. 
%with a half-wave plate.
We made pairs of measurements at $P$ and $P/2$ while keeping the probe duration, the magnetic field and all other parameters of the sequence the same. This came at the cost of using a non-optimal probe pulse in both cases, one stronger and one weaker than the optimum. 
%\czedit{This was done at the expense of using, in both cases, non-optimal probe pulses, one stronger and one weaker than the $\pi$ pulse, i.e., the pulse that transfers the entire population from the ground to the excited state.} 
We also made measurements with optimal pulses. Overall we explored a power range from 1 to 500~µW, corresponding to a central beam intensity $I_0$ experienced by the atoms ranging from $15~\text{mW/cm}^{2}$ to $7.9~\text{W/cm}^{2}$. 
%The corresponding maximum clock light shift is $\sim 16$~Hz.

Once again, the probe intensity exhibits significant inhomogeneities across the atomic sample that must be taken into account. To do so, we expand the simple model used for extracting $\alpha$ to include the transverse distribution of detunings caused by the probe light shift. For optimal pulses, the model shows that these inhomogeneities reduce the effective clock shift to about $0.76\times \kappa I_0$. For $P$–$P/2$ measurements, the predicted slope is smaller by an additional factor of $0.9$, and their linear extrapolation to $P=0$ retains a bias of $\sim 0.06$ times the shift at $P$. Experimentally, we measure slopes of $33$~mHz/µW for optimal pulses with $P$ in the $10$–100~µW range. For the $P$–$P/2$ method, we obtain slopes from $11$ to $29$~mHz/µW for $P$ in the $50$–$500$~µW range, with a mean of $22$~mHz/µW. This is not fully consistent with the predicted ratio of slopes between the two methods.
%The model indicates that, for optimal pulses, the clock shift is $\sim 0.76\times\kappa I_0$. It also predicts that the slope for $P$-$P/2$ measurements is smaller by an additional factor of $0.9$, and that their linear extrapolation to $P=0$ retains a bias of $\sim 0.06$ times the shift at $P$. Experimentally, we measured slopes of $33$~mHz/$\mu$W for optimal pulses with $P$ in the $10-100~\mu$W range. $P$-$P/2$ measurements give slopes of $11-29$~mHz$/\mu$W for $P$ in the $50-500~\mu$W range, with a mean value of $22$~mHz$/\mu$W. 
We further consider that the beam shape could significantly deviate from a Gaussian profile and that its alignment with the atomic sample may be imperfect or fluctuate over time. As a result, $20\%$ is a reasonable relative uncertainty for our determination of $\kappa$ given in Table~\ref{fig_table_rabi}, which is derived from measurements with the optimal pulses.

\begin{table}[h]
    \caption{\label{fig_table_rabi}%
    Theoretical and experimental (this work) values along with their uncertainties for the coupling coefficient, the QZS coefficient, and the probe light shift coefficient. We note that for $\kappa$, due to the simplified model, we do not expect an agreement between the theoretical and experimental values.}
    \begin{ruledtabular}
    \begin{tabular}{c c c c}
        & $\alpha/2\pi$  & $\beta/2\pi$  & $\kappa/2\pi$  \\ 
        & $\frac{\text{Hz}}{\text{T}\sqrt{\text{mW/cm}^2}}$ & $\frac{\text{MHz}}{\text{T}^2}$ & $\frac{\text{mHz}}{\text{mW/cm}^2}$ \\ 
        \colrule
        \text{Basic theory}  & 54%61
        & -2.465  & 0.38  \\ 
        \text{Experiments}  & 45 $\pm$ 9  & -2.969 $\pm$ 0.16
        %-2.969 $\pm$ 0.16
        & 2.74 $\pm$ 0.55 \\
        %2.80 $\pm$ 0.56  \\ 
        %& 1.21 $\pm$ 0.2  \\ 
    \end{tabular}
    \end{ruledtabular}
\end{table}

We now aim to determine the factor of merit $\xi$ for the induced mixing scheme, as defined in \cite{Taichenachev2008}. It characterizes the relationship between the coupling strength, which we seek to maximize, to the resulting field-induced shifts, which are undesirable. Specifically, the Rabi angular frequency can be expressed in terms of the probe light shift $\Delta_L$ and the quadratic Zeeman shift $\Delta_B$ as $\Omega_{12}~=~\alpha \sqrt{|\frac{\Delta_L}{\kappa} \, \frac{\Delta_B}{\beta}|}~=~\xi \sqrt{|\Delta_L \Delta_B|}$, where $\xi = \alpha/\sqrt{|\kappa \beta|}$ represents in other terms the effective excitation strength relative to the induced shifts. Table~\ref{fig_table_rabi} lists the experimentally determined coefficients $\alpha$, $\beta$ and $\kappa$.

%We now expand the discussion to account for the detailed level structure of a real atom, by recalculating the Rabi angular frequency in terms of the probe light shift $\Delta_L$ and the quadratic Zeeman shift $\Delta_B$, following the method proposed in \cite{Taichenachev2008}. We rewrite the Rabi angular frequency $\Omega_{12} = \alpha \sqrt{\left|\frac{\Delta_L}{\kappa} \, \frac{\Delta_B}{\beta}\right|} = \xi \sqrt{|\Delta_L \Delta_B|}$, with $\xi$ is a factor of merit that describes the excitation strength relative to the magnitudes of the induced field shifts. By definition, $\xi = \frac{\alpha}{\sqrt{\kappa \beta}}$. Table \ref{fig_table_rabi} summarizes the experimentally obtained values for the QZS coefficient, the magnetic field induced coefficient, and the probe light shift coefficient.

Table~\ref{fig_table_rabi} also gives theoretical values that we computed taking into account the dominant coupling of ${}^{3}\text{P}_{0}$ to ${}^{3}\text{P}_{1}$, following \cite{Taichenachev2008}, and adding the effect of ${}^{3}\text{P}_{0}$ to ${}^{1}\text{P}_{1}$. 
For the induced coupling $\alpha$, we find that ${}^{1}\text{P}_{1}$ accounts for about 10\% of the total effect. For the QZS coefficient $\beta$, the influence of ${}^{1}\text{P}_{1}$ is found negligible (about 0.1$\%$).
%of the order of  $10^{-3}\times\beta$
Due to the large experimental uncertainty in $\alpha$, mainly arising from probe beam inhomogeneities, we retain the theoretical value of $\alpha$ to estimate $\xi$. On the contrary, for $\kappa$, we retain our experimental value because we know that the simplified model cannot accurately account for the polarizabilities of the clock states $^{1}\text{S}_{0}$ and ${}^{3}\text{P}_{0}$ and thereby not capture the probe light shift. With this approach, we obtain our best estimate of the factor of merit $\xi$, given in Table~\ref{tab_xi_values}, for the bosonic transition in Hg. It stands as the most favorable species, together with Yb, among several other considered species.

\begin{table}[h]
    \caption{\label{tab_xi_values}%
    Factor of merit for several bosonic species considered for optical lattice clocks \cite{Taichenachev2008}, including ${}^{198}\text{Hg}$ (this work).}
    \begin{ruledtabular}
    \begin{tabular}{c c c c c c}
        & ${}^{198}\text{Hg}$ & \text{Sr} & \text{Yb} & \text{Ca} & \text{Mg} \\ \colrule
        \text{$\xi$} & 0.60% 0.59 with new PLS, 0.89 of old QZS, 0.98 if QZS theo!
        & 0.30 & 0.60 & 0.28 & 0.28 \\
    \end{tabular}
    \end{ruledtabular}
\end{table}

\begin{table}
    \caption{\label{table_198_budget}%
    Typical systematic effect corrections and uncertainties of the ${}^{198}\text{Hg}$ clock during the campaign for determining the ${}^{198}\text{Hg}/{}^{87}\text{Sr}$ optical frequency ratio. The QZS %\sout{correction and uncertainty} 
    is given for a current of $3.6$~A (field $\sim 5.5$~mT), and the probe light shift %\sout{correction and uncertainty} 
    for a power of 100 µW ($I_0\sim 1.6$~W/cm$^2$). During the campaign, data were acquired at these values or lower.}
    %\czedit{The dominant contributions to the total uncertainty arise from the probe light shift on one hand, and from the cold collisional shift and the linear lattice light shift on the other. The latter two are currently limited by statistical uncertainties and could be further reduced with additional measurements.\sbedit{update uncertainties QZS, PLS,...}}
    
    \begin{ruledtabular}
    \begin{tabular}{l c c}
        \textbf{Effect} & \makecell{\textbf{Correction} \\ \textbf{($\times 10^{-17}$)}} & \makecell{\textbf{Uncertainty} \\ \textbf{($\times 10^{-17}$)}} \\
        \colrule
        Quadratic Zeeman effect       & 7868.7 & 4.9  \\
        Magnetic field fluctuations   & 0 & 2.9 \\
        Residual magnetic field       & 0 & 4.3\\
        Cold collisional shift        & 4.1 & 17.3 \\
        Linear lattice light shift    & 0    & 32.3 \\
        Nonlinear lattice light shift & 2.1    & 2.1  \\
        Blackbody radiation          & 14.6   & 1.5  \\
        Background gas collisions     & 3.0    & 3.0  \\
        Probe light shift              & -291.8 & 58.4 \\
        %Probe light shift instability & 0  & ? \\
        \colrule
        \textbf{Total} & \textbf{7600.6} & \textbf{69.4} \\
    \end{tabular}
    \end{ruledtabular}
\end{table} 
    
%We operated the mercury bosonic lattice clock during two comparison campaigns against the local strontium clocks. The first campaign was preliminary (Jun.~2024), while the second was performed after substantial operational optimization (Sep.~2025). In total, we accumulated $\sim 40$~hours, including the measurements reported above, in particular those of the probe light shift. Measurements in configurations deemed favorable, i.e., probe power $\leq 100~\mu$W and current $\leq 3.6$~A, were used to determine the $^{198}\text{Hg}/^{87}\text{Sr}$ optical frequency ratio. The corresponding systematic uncertainty budget is given in Table~\ref{table_198_budget}.
We operated the bosonic mercury lattice clock in two comparison campaigns against the local strontium clocks: a preliminary run in Jun.~2024 and a second after major operational optimization in Sep.~2025. In total, we accumulated about 40~hours of data. The systematic uncertainty budget is shown in Table~\ref{table_198_budget}.
The blackbody radiation values rely on theoretical polarizabilities given in \cite{Hachisu2008}. Background gas collision values rely on \cite{Gibble2013} and its practical implementation \cite{Guidelines2019}, applied to Hg \cite{Clara_PhD}. The $^{87}\text{Sr}$ counterparts have negligible uncertainties of $4.6\times 10^{-17}$.
We obtain a frequency ratio of  $^{198}\text{Hg}/^{87}\text{Sr}~=~2.629~315~734~684~118~1$ with a fractional uncertainty of $6.9\times 10^{-16}$. From this measured ratio and the CCTF 2021 recommended frequency and recommended uncertainty for ${}^{87}\text{Sr}$ \cite{Margolis2024}, we derive the frequency of the bosonic ${}^{198}\text{Hg}$ clock transition: $\nu_{\text{${}^{198}\text{Hg}$}}~=~1~128~575~945~288~666.3\pm 0.78 $~Hz.

The probe light shift dominates the uncertainty budget (Table~\ref{table_198_budget}), mainly due to probe beam inhomogeneities and the difficulty of accurately controlling the probe intensity. This uncertainty could be further reduced using a hyper-Ramsey interrogation scheme, which effectively cancels field shifts and their related uncertainties \cite{Taichenachev2010, ZanonWillette2018}. This method has been shown to suppress probe shifts by factors up to $2\times10^3$ \cite{Hobson2015} and even $10^4$ \cite{Huntemann2012}.
Once in place, it will become possible to fully exploit the advantages of bosonic isotopes of Hg: a sensitivity to blackbody radiation 30 and 16 times smaller than Sr and Yb respectively, associated to an unlimited $^3$P${_0}$ state lifetime to take the full benefit of best present and future ultrastable lasers. Highly accurate frequency ratio measurements involving Hg can contribute to validating optical clocks in view of a redefinition of the second \cite{Dimarcq2024} and to fundamental physics tests relying on diverse optical transitions \cite{Roberts2020a,Safronova2018,Uzan2025}.
  
%\textit{(Optional : Precise characterization of new atomic transitions is also essential for the ongoing effort to redefine the SI second \cite{Dimarcq2024, Margolis2024}, to exploit the full potential of optical clocks, and to enable new tests of fundamental physics [ref?].)}}

%acknowledgments
Acknowledgments: Within the french national metrology network coordinated by LNE, LNE-OP at LTE-Observatoire de Paris is the national metrology institute for time, frequency and gravimetry. We acknowledge contributions of LTE electronic, mechanic and IT support services. C. Zyskind received funding from Sorbonne Université (PhD grant CMA FQPS n°ANR-21-CMAQ-0001 dans le cadre de France 2030).


%\nocite{*}s

\bibliography{apssamp}% Produces the bibliography via BibTeX. 

\end{document}